\documentclass{LT23auth}
\usepackage{graphicx}

\newcommand{\lsco} {La$_{2-x}$Sr$_x$CuO$_4$}

\newcommand{\prb} {{Phys. Rev. B}}
\newcommand{\prl} {{Phys. Rev. Lett.}}
\newcommand{\jpsj} {{J. Phys. Soc. Jpn.}}

\begin{document}

\begin{frontmatter}

\title{Magnetic order and superconductivity in \lsco: a review}

\author[address1]{M.-H. Julien\thanksref{thank1}}

\address[address1]{Laboratoire de Spectrom\'etrie Physique, Universit\'e J. Fourier Grenoble
1, BP87, 38402 Saint Martin d'H\`eres, France }

\thanks[thank1]{Corresponding author.  E-mail: Marc-Henri.Julien@ujf-grenoble.fr}

\begin{abstract}
High-$T_c$ copper oxides of the \lsco~family show a very clear
case of {\it competition} between antiferromagnetic (AF) order and
superconductivity. Magnetic order can, however, {\it coexist} with
superconductivity, and the experimental evidence for frozen
magnetic moments in superconducting samples is reviewed here. The
primary characteristics of the magnetic order are summarized and
some open questions are outlined, particularly concerning the
intrinsic or extrinsic nature of this order around $x=0.12$.
\end{abstract}

\begin{keyword}
high-$T_c$ superconductivity; magnetic order; stripes; magnetic
resonance
\end{keyword}
\end{frontmatter}

\begin{figure*}
\begin{center}\leavevmode
\includegraphics[width=0.6\linewidth]{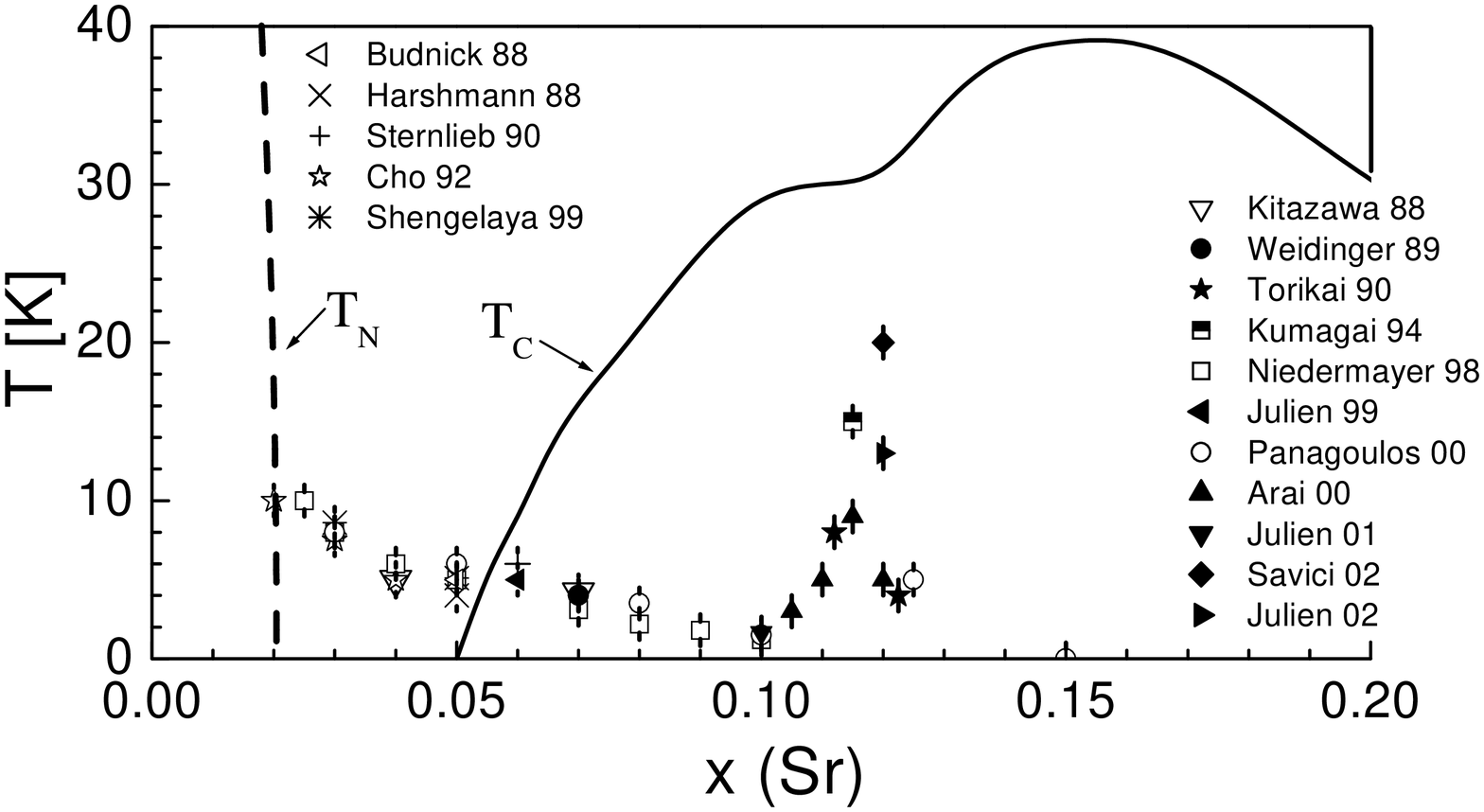}
\vspace{-2.3cm} \caption{Magnetic phase diagram of \lsco: data
points correspond to the temperature of magnetic freezing, $T_g$,
inferred from $\mu$SR, NQR and NMR measurements (references are
given in the bibliography). A systematic error $\Delta T_g
=\pm1$~K was chosen. Note that for Ref.~\cite{Kitazawa88} a
concentration of $x=0.07$ was used for the plot (0.08 in the
original paper) since the sample has a $T_c$ of 11~K. Other
reports of a frozen magnetic state, or of strong slowing down of
spin fluctuations, do not appear here. For example, Ohsugi
\etal~observed a broadening of the $^{139}$La NQR lines due to the
static internal magnetic field, but no freezing temperature was
deduced~\cite{Ohsugi94}. Data from Fe EPR and Fe M\"ossbauer
spectroscopy, which use a dilute impurity in the CuO$_2$ planes to
probe the magnetic properties, are also omitted here (the role of
impurities would require additional discussion). Other EPR data
can be found in Refs.~\cite{Kataev93,Kochelaev97}. Data on
marginal samples (anomalous $T_c$~\cite{Lin00}) or with
questionable criteria for $T_g$, or isolated experiments from a
single technique~\cite{Gutsmiedl87} are not discussed. Another
material that is omitted here is La$_2$CuO$_{4+\delta}$, which
also shows superconductivity and magnetic order, but with some
differences with respect to \lsco.} \label{phasediagram}
\end{center}
\end{figure*}

\section{Introduction}

In 1988, R.B. Laughlin wrote about high $T_c$ superconductors:
"The systems in question are inherently magnetic. Stoichiometric
La$_2$CuO$_4$ is an ordered spin-1/2 antiferromagnet and an
insulator. Doping the material by substituting Sr for about 3\% of
the La destroys the magnetic order [...]. It is hard to understand
how doping at this level could have destroyed all the spins. A
more reasonable guess is that the extra holes make ordering more
difficult, and that the spins are still present in some sort of
'quantum spin liquid' state"~\cite{Laughlin88}.

For a number of years thereafter, experimental studies failed to
establish a consensus on whether localized spins indeed survived
in metallic samples, possibly in a kind of spin-liquid state, or
whether correlation effects in a metal were sufficient to explain
the observed magnetic fluctuations. Despite much evidence that the
superconducting phase retains some memory of the AF
one~\cite{dopedAF}, some considered that tangible signs of
localized spins were lacking for superconducting concentrations
($0.06\leq x \leq 0.28$ in \lsco): N\'eel order was rather far
away ($x<0.02$) and not much attention was paid to the frozen
magnetic state for $0.02<x<0.06$.

Understanding of this issue has evolved since the discovery of
charge stripes in
La$_{1.48}$Nd$_{0.4}$Sr$_{0.12}$CuO$_4$~\cite{Tranquada95} and
related compounds~\cite{remLTT}. In this class of materials,
self-organization of the doped holes into linear ribbons leads to
{\it long range magnetic order} in the hole-depleted regions, even
for hole doping close to the optimal value for superconductivity
($\sim$15 \%). This spectacular "reappearance" of the spins was
known before Tranquada's experiment, but the discovery of stripes
renewed interest in magnetic order, and contributed to the view of
high-$T_c$ materials as doped antiferromagnets~\cite{impurities}.

This paper is a short review on magnetic order in
\lsco~superconductors. These represent a particularly interesting
case, because they lie in between the LTT materials~\cite{remLTT},
where superconductivity is severely suppressed, and higher $T_c$
systems such as YBCO where magnetic order is less obvious.

\section{Reviewing literature}

There are many reports of magnetic order coexisting with
superconductivity in \lsco, starting with 
Kitazawa \etal~in 1988~\cite{Kitazawa88}. These results did not
attract significant attention mostly because doubts were raised
concerning the homogeneity of the samples, and because \lsco~was
considered as an atypical member of the high-$T_c$ family. This
issue was reopened around
1998~\cite{Niedermayer98,Julien99,Suzuki98} and most of the
previous results were confirmed quantitatively.

Figure~1 shows most of the data available in the literature (to
the author's knowledge), for the magnetic transition temperature
$T_g$ obtained from muon spin rotation ($\mu$SR), nuclear magnetic
resonance (NMR) and nuclear quadrupole resonance (NQR) in
superconducting
~\cite{Kitazawa88,Niedermayer98,Julien99,Weidinger89,Uemura89,Torikai90,Kumagai94,Panagopoulos00,Arai00,Julien01,Savici02,Julien02}
and non-superconducting samples~\cite{musrSG,Cho92}. The reason
for this selective presentation of the magnetic resonance data is
twofold: numerous studies can be found in the literature which are
lower-energy probes than neutron scattering (NS). Because the NS
signal is quasi-elastic rather than purely elastic (the
integration window is usually not less than 0.5meV), and freezing
of the moments is gradual in these compounds, the apparent onset
of magnetic order occurs at higher $T$ for NS than for magnetic
resonance. NS studies of magnetic order in superconducting
\lsco~can be found in
Refs.~\cite{Suzuki98,Kimura99,Kimura00,Katano00,NeutronSG,Wakimoto01,Fujita02,Lake02}.
On the other hand, magnetic resonance techniques can be considered
as true low-energy probes: in the non-superconducting phase
($0.02\leq x \leq 0.05$), the transition temperature $T_g$,
defined at their time scale, is indeed very close (typically 1~K)
to the $T_g$ inferred from SQUID measurements~\cite{squidSG},
which are almost static.

\section{Magnetic order in the phase diagram}

The main features of the magnetic phase diagram on Fig.~1 can be
summarized as follows:

$\bullet$ The agreement between the data is good, and there is no
doubt that the magnetic phase for $0.02\leq x \leq 0.05$ continues
far into the superconducting region. Magnetic order thus coexists
with superconductivity in underdoped \lsco.

$\bullet$ Except at the point where the $T_c~vs.~x$ and the
$T_g~vs.~x$ lines cross, nowhere in the phase diagram does the
magnetic transition coincide with the onset for superconductivity
(see the discussion in~\cite{Julien01} for the explanation of
earlier confusion on this point).

$\bullet$ Magnetic order seems to persist at very low $T$ for
$x=0.15$~\cite{Weidinger89,Panagopoulos00,Ohsugi94}, and up to
$x=0.19$~\cite{Panagopoulos02}, although there is not full
agreement on this issue~\cite{Kiefl89}.

$\bullet$ Some scatter in the data can be seen around $x=0.12$.
For example, a $\mu$SR study detects the appearance of frozen
moments near 20~K in
La$_{1.88}$Sr$_{0.12}$CuO$_4$~\cite{Savici02}, while an NMR study
in a very similar single crystal, defines $T_g=13$~K as the $T$ at
which the {\it average} relaxation rate $1/T_1$ is
maximum~\cite{Julien02}. If the volume fraction of magnetic order
grows on cooling down (distribution of $T_g$ values), the
discrepancy between the criteria is not surprising. In addition,
the strong $x$ dependence of $T_g$ around $x=0.12$ may contribute
to the scatter between data from different samples, and the
somewhat higher energy scale of $\mu$SR with respect to NMR may
become noticeable since $T_g$ is higher.

$\bullet$ In any event, there is a clear {\it enhancement of $T_g$
around $x=0.12$}, coinciding with a slight suppression of $T_c$.
This suggests the same '1/8 anomaly' as in LTT species, although
with the following differences: Superconductivity is only weakly
affected here (note however that a few \lsco~samples around
$x=0.12$ have an anomalously low $T_c$, which is yet to be
understood~\cite{Katano00,Lin00,Goto94}). The maximum of $T_g$
seems to be occur around $x=0.115$-0.12 rather than at
$0.125=1/8$.

$\bullet$ The overall magnetic phase diagram in Fig.~1 is similar
to that of LTT
La$_{1.8-x}$Eu$_{0.2}$Sr$_x$CuO$_4$~\cite{Klauss00}, but the peak
of $T_g$ around $x=0.12$ is much narrower in \lsco. This makes a
clearer distinction between the behavior close to $x=0.12$, and
the monotonic decrease of $T_g~vs.~x$ for $x\leq 0.10$.

\section{Other features of magnetic order}

$\bullet$ For $0.02\leq x\leq 0.05$, characteristic features of
spin-glasses (in the loose sense: these features are also seen in
diluted antiferromagnets) are observed in the bulk
magnetization~\cite{squidSG}. Because of the Meissner effect, no
such study could be performed in superconducting samples. However,
the continuous decrease of $T_g$ from $x=0.02$ up to $x=0.10$
suggests that the magnetic state is similar on both sides of the
non-superconductor to superconductor transition. Furthermore,
$\mu$SR~\cite{Niedermayer98,Weidinger89,Panagopoulos00} and
NMR~\cite{Julien99,Julien01,Singer02} show that sizeable disorder
(spatial inhomogeneity) in both the spin dynamics and the static
local magnetization also exists for $x\geq0.06$. The slowing down
of magnetic fluctuations is also similar above and below $x=0.06$,
and is more gradual than for a conventional 3D N\'eel transition.
Magnetic order is thus considered to have some glassy character in
superconducting samples, and the transition temperature is usually
called $T_g$.

$\bullet$ The frozen magnetic state between $x=0.02$ and $x=0.10$
has been named a 'cluster spin-glass', in order to reconcile the
glassy features with the existence of domains of staggered
magnetization~\cite{Julien99,Cho92,Emery}. It has also been clear
that some kind of charge segregation/order is necessary in order
to explain the existence of frozen AF clusters at concentrations
as high as $\sim$12\%. However, it was recently shown that the
frozen state could actually be described as diagonal (with respect
to Cu-O bonds) stripes for $x\leq0.05$~\cite{NeutronSG}, within
the neutron scattering time window, collinear stripes for $x\geq
0.06$~\cite{Suzuki98,Kimura99,Kimura00}, and the coexistence of
both around $x=0.06$~\cite{Fujita02}. The existence of magnetic
stripes explains the AF domains of the cluster spin-glass, but it
does not allow one to deduce whether all of the spatial
inhomogeneity is due to the stripe pattern or whether additional
phenomena, such as phase separation between striped and
non-striped regions, take place.

$\bullet$ The magnetic order in question is not magnetic-field
induced. Lake \etal~observe an enhancement of the Bragg peak
intensity by applying a field, and a change in its $T$ dependence,
but magnetic order at $x=0.10$ is known to exist even in zero
field both from measurements by the same authors~\cite{Lake02},
and from earlier
studies~\cite{Niedermayer98,Panagopoulos00,Julien01,Kimura99,Ohsugi94}
(see also~\cite{Katano00} for $x=0.12$). Note that the $T$ at
which a neutron diffraction signal appears does not seem to vary
with the field~\cite{Lake02}, and that no strong modification of
slowing-down with the field could be detected in
NMR~\cite{Julien99,Julien01}.

$\bullet$ The magnetic correlation length at low $T$ is shortest
around $x=0.06$ ($\xi\sim$~20~\AA) and has a strong peak around
$x=0.12$ ($\xi\geq$~200~\AA)~\cite{Wakimoto01,Fujita02}, where
stripes are actually slanted~\cite{Kimura00}. The ordered moment
decreases with $x$ up to $x\simeq 0.10$ but it seems to be
enhanced at
$x=0.12$~\cite{Niedermayer98,Kumagai94,Savici02,Wakimoto01}.

\section{Is magnetic order intrinsic ?}

A recent $\mu$SR study for $x=0.12$ indicates that the frozen
magnetic regions represent not more that 18 \% of the sample
volume, at low$T$~\cite{Savici02}. This immediately raises the
question of the intrinsic character of the magnetic
phase~\cite{remNMR}.

Savici \etal~propose that there is phase separation between
regions with (striped) magnetic order and superconducting regions
without magnetic order~\cite{Savici02}. Alternative and/or
complementary explanations should however be considered. First,
there is evidence that local LTT distortions exist in the LTO
phase~\cite{remLTT} of
\lsco~\cite{Moodenbaugh98,Bozin99,Horibe00}. As noted in
Refs.~\cite{Moodenbaugh98,Tranquada98}, this should cause local
pinning of stripes, and thus nucleate magnetic order. The fact
that $T_c$ is only weakly affected in
La$_{1.88}$Sr$_{0.12}$CuO$_4$ and the absence of a 1/8-anomaly for
$T_g$ in a Y-doped Bi2212~\cite{Panagopoulos02}, a material which
does not show the LTT instability, could support this hypothesis.
Another possibility is that the magnetic fraction at the $\mu$SR
time scale is reduced by rapid spin flips. These spin flips might
be produced by transverse stripe fluctuations~\cite{stripefluct}.
It is instructive to remark that in LTT species, where stripes are
supposedly more static, there is already evidence for substantial
averaging of the local magnetization at the NMR time
scale~\cite{Hunt01}; a magnetic volume fraction of 100\% is
observed at $x=0.12$, but it is reduced by {\it half} at
$x=0.15$~\cite{Kojima00}. At present it is thus unclear whether
the $T_g$ anomaly and the reduced magnetic fraction in $\mu$SR for
$x=1/8$ in~\lsco~are intrinsic or if they are related to extra
pinning by lattice distortions in some parts of the sample.

For $0.02<x\leq 0.10$, magnetic order seems to be intrinsic as a
large fraction, if not all, of the muons see an internal field,
according to
Refs.~\cite{Kitazawa88,Niedermayer98,Weidinger89,Torikai90}.

All studies to date and the good agreement between data sets from
many different samples, point to the existence of bulk
superconductivity in these systems, except close to the onset
value $x\simeq0.06$ and possibly at $x=0.12$ (this case is not
clear)~\cite{Niedermayer98,Torikai90,Savici02,SC}. Note that these
studies ascertain that the samples studied by NMR or $\mu$SR are
representative of superconducting \lsco, but they do not rule out
the existence of normal regions and/or unpaired carriers.

In conclusion, coexistence of magnetic order with
superconductivity is intrinsic in a significant part of the phase
diagram of \lsco. Since these are bulk phases, without any hint of
macroscopic phase separation, the coexistence has to occur on a
small length scale. Nevertheless, the data discussed here do not
seem to provide firm answers to crucial
questions~\cite{otherfreezing}: Is the stripe picture able to
account alone for the coexistence~? Is there a strong
hybridization between the magnetic and the superconducting
entities or should they be considered as distinct phases~? More
generally, which ingredients make the coexistence possible :
spatial segregation, different orbitals (Cu and O), different
energy scales, etc.~?

\begin{ack}
B. Normand, V. Mitrovi\'c and C. Berthier are thanked for
discussions and reading of the manuscript.
\end{ack}


\begin{thebibliography}{9}
\bibitem{Laughlin88} R. B. Laughlin, Science {\bf  242} (1988) 525.

\bibitem{dopedAF} For instance in \lsco: D. C. Johnston, \prl~{\bf
62} (1989) 957; T. Imai \etal, \prl~{\bf 70} (1993) 1002; P.
Carretta \etal, Eur. phys. J. B {\bf 10} (1999) 233; K. Yamada
\etal, \jpsj~{\bf 64} (1995) 2742.

\bibitem{Tranquada95} J. M. Tranquada \etal, Nature {\bf 375} (1995) 561.

\bibitem{remLTT} The main difference between "regular" \lsco~and
La$_{2-x-y}$(Nd,Eu)$_{y}$Sr$_{x}$CuO$_4$ or
La$_{2-x}$Ba$_{x}$CuO$_4$, is that the crystallographic structure
of the former is Orthorhombic at Low $T$ (LTO) in most of the
relevant doping range, while it is Tetragonal (LTT phase) in the
latter. The two phases correspond to two directions of the
planar-oxygen buckling pattern, which differ by 45$^\circ$.

\bibitem{impurities} It should be noted that the physics of non-magnetic
impurities also provides a striking example of the reappearance of
magnetic moments.

\bibitem{Kitazawa88}
H. Kitazawa \etal, Solid State Commun. {\bf 67} (1988) 1191.

\bibitem{Niedermayer98}
Ch. Niedermayer \etal, Phys. Rev. Lett. {\bf 80} (1998) 3843.

\bibitem{Julien99} M.-H. Julien \etal, Phys. Rev. Lett. {\bf 83} (1999) 604;
Appl. Magn. Res. {\bf 3-4} (2000) 287 (cond-mat/9909351).

\bibitem{Suzuki98}
T. Suzuki \etal, \prb~ {\bf 57} (1998) R3229.

\bibitem{Weidinger89}
A. Weidinger \etal, Phys. Rev. Lett {\bf 62} (1989) 102; Phys.
Rev. Lett. {\bf 63} (1989) 2539; Phys. Rev. Lett. {\bf 63} (1989)
1188.

\bibitem{Uemura89} Y. J. Uemura \etal, Phys. Rev. Lett. {\bf 62} (1989) 2317.

\bibitem{Torikai90} E. Torikai \etal, Hyp. Int. {\bf 63} (1990) 271.

\bibitem{Kumagai94} K. Kumagai \etal, Hyp. Int. {\bf 86} (1994) 473.

\bibitem{Panagopoulos00} C. Panagopoulos \etal, Physica C, {\bf 341-348} (2000) 843.

\bibitem{Arai00} J. Arai \etal, Physica B {\bf 289-290}
(2000) 347.

\bibitem{Julien01} M.-H. Julien \etal, Phys. Rev. B {\bf 33} (2001) 144508.

\bibitem{Savici02} A. T. Savici \etal, \prb~{\bf 66} (2002) 014524.

\bibitem{Julien02} M.-H. Julien \etal, unpublished.

\bibitem{musrSG} J. I. Budnick \etal, Europhys. Lett. {\bf 5} (1988) 651; D. R. Harshman \etal, Phys.
Rev. B {\bf 38} (1988) 852; B. J. Sternlieb \etal, Phys. Rev. B
{\bf 41} (1990) 8866; A. Shengelaya \etal, \prl {\bf 83} (1999)
5142.

\bibitem{Cho92} J. H. Cho \etal, Phys. Rev. B {\bf 46} (1992) 3179.

\bibitem{Kimura99} H. Kimura \etal, Phys. Rev. B {\bf 59} (1999) 6517.

\bibitem{Kimura00} H. Kimura \etal, \prb~{\bf 61} (2000) 14366.

\bibitem{Katano00} S. Katano \etal, \prb~{\bf 62} (2000) R14677.

\bibitem{NeutronSG} S. Wakimoto \etal \prb~{\bf 60} (1999) R769; \prb~{\bf 61} (2000) 3699;
M. Matsuda \etal, \prb~{\bf 62} (2000) 9148.

\bibitem{Wakimoto01} S. Wakimoto \etal, \prb~ {\bf 63} (2001) 172501.

\bibitem{Fujita02} M. Fujita \etal, \prb~{\bf 65} (2002) 064505.

\bibitem{Lake02} B. Lake \etal, Nature {\bf 415} (2002) 299.

\bibitem{squidSG} M. E. Filipkowski \etal, Physica C {\bf 167} (1990) 35; F.C. Chou
\etal, Phys. Rev. Lett. {\bf 75} (1995) 2204; S. Wakimoto \etal,
\prb~{\bf 62} (2000) 3547; A. N. Lavrov \etal, \prl~{\bf 87}
(2001) 017007.

\bibitem{Ohsugi94} S. Ohsugi \etal, J. Phys. Soc. Jpn. {\bf 63} (1994) 2057.

\bibitem{Kataev93} V. Kataev \etal, \prb~ {\bf 48} (1993) 13042.

\bibitem{Kochelaev97} B. I. Kochelaev \etal, \prl~{\bf 79} (1997) 4274.

\bibitem{Lin00} Y. Lin \etal, \prb~ {\bf 61} (2000) 7130.

\bibitem{Gutsmiedl87} P. Gutsmiedl \etal, \prb~{\bf 36} (1987) 4043.

\bibitem{Panagopoulos02} C. Panagopoulos \etal, \prb~ {\bf 66} (2002) 064501.

\bibitem{Kiefl89} R. F. Kiefl \etal, Phys. Rev. Lett. {\bf 63} (1989) 2136.

\bibitem{Goto94} T. Goto \etal, \jpsj {\bf 63} (1994) 3494.

\bibitem{Klauss00} H. H. Klauss \etal, \prl {\bf 85} (2000) 4590.

\bibitem{Singer02} P. M. Singer \etal, \prl {\bf 88} (2002) 047602.

\bibitem{Emery} V. J. Emery, Hyp. Int. {\bf 63} (1990) 13;
V. J. Emery and S. A. Kivelson, Physica {\bf 209C} (1993) 597; R.
J. Gooding \etal, \prb~{\bf 55} (1997) 6360.

\bibitem{remNMR} This result might appear to contradict the absence of
any NMR signal with a distinct, longer, $T_1$ that could be
expected from sizeable non-frozen
regions~\cite{Julien01,Julien02}. However, NMR cannot exclude that
such an NMR signal is hidden by spin diffusion processes which
homogeneizes the $T_1$ of the nuclei outside frozen areas with
that of nuclei inside.

\bibitem{Moodenbaugh98} A. R. Moodenbaugh \etal, \prb~{\bf 58} (1998) 9549.

\bibitem{Bozin99} E. S. Bozin \etal, \prb~ {\bf
59}, (1999) 4445.

\bibitem{Horibe00} Y. Horibe \etal, \prb~{\bf 61} (2000) 11922.

\bibitem{Tranquada98} J. Tranquada, in {\it Neutron Scattering in
Layered Copper-Oxide Superconductors}, Edited by A. Furrer (Kluwer
Academic, Dordrecht, 1998), p. 225.

\bibitem{Hunt01} A. W. Hunt \etal, \prb~{\bf 64} (2001) 134525.

\bibitem{Kojima00} K. M. Kojima \etal, Physica B {\bf 289-290}
(2000) 343.

\bibitem{stripefluct} Oscillation of an O-centered stripe across a Cu site averages the
magnetic moment on this site to zero, on a time scale which is
long compared to the jump frequency. This moment is only reduced
by a factor 2 for Cu-centered stripes. I thank S. Kivelson for
pointing out this distinction.

\bibitem{SC} For example T. Nagano \etal,
\prb~{\bf 48} (1993) 9689; T. Suzuki \etal, \prb {\bf 60} (1999)
10500; Y. M. Huh \etal, \prb~{\bf 63} (2000) 064512.

\bibitem{otherfreezing} It is interesting to compare these data with recent
works by T. Sasagawa \etal, cond-mat/0208014
(La$_2$Cu$_{1-x}$Li$_x$O$_4$) and A. Kanigel \etal, \prl~{\bf 88}
(2002) 137003.

\end{thebibliography}
\end{document}